\documentclass[journal,onecolumn]{IEEEtran}
\IEEEoverridecommandlockouts
\usepackage{cite}
\usepackage{float}
\usepackage{amsmath,amssymb,amsfonts}
\usepackage{algorithmic}
\usepackage{graphicx}
\usepackage{textcomp}
\usepackage{comment}
\usepackage{xcolor}
\usepackage{booktabs}
\usepackage{siunitx}
\usepackage{tikz}
\usetikzlibrary{shapes, arrows.meta, positioning}
\usepackage{tabu}
\usepackage{multicol}
\usepackage{multirow}
\usepackage{subcaption}
\usepackage{hyperref}
\usepackage{xurl}
\usepackage{xcolor,colortbl}
\usepackage[flushleft]{threeparttable}
\usepackage{tikz}
\usepackage{multirow} 
\usepackage[utf8]{inputenc}
\usepackage[linesnumbered, ruled, vlined]{algorithm2e}

\hypersetup{
    colorlinks=true,
    linkcolor=black,
    urlcolor=black,
    filecolor=black,
    citecolor=black,
    linkbordercolor=white, 
    pdfborderstyle={/S/U/W 1}, 
}
\usepackage[flushleft]{threeparttable}
\def\BibTeX{{\rm B\kern-.05em{\sc i\kern-.025em b}\kern-.08em
    T\kern-.1667em\lower.7ex\hbox{E}\kern-.125emX}}

\begin{document}

\newcommand{\pink}[1]{\textcolor[rgb]{0.00,0.00,1.00}{{}}}   
\newcommand{\CLLL}{\rowcolor{blue!25}}

\title{Alleviating Attack Data Scarcity: SCANIA's Experience Towards Enhancing In-Vehicle Cyber Security Measures} 

\author{

\IEEEauthorblockN{Frida Sundfeldt\IEEEauthorrefmark{1}\IEEEauthorrefmark{2}, Bianca Widstam\IEEEauthorrefmark{1}\IEEEauthorrefmark{2}, Mahshid Helali Moghadam\IEEEauthorrefmark{1}, Kuo-Yun Liang\IEEEauthorrefmark{1}, and Anders Vesterberg\IEEEauthorrefmark{1} }

\IEEEauthorblockA{\IEEEauthorrefmark{1}
Cloud and Embedded Platform, Traton AB, Södertälje, Sweden\\
\{mahshid.helali.moghadam, kuo-yun.liang, and anders.vesterberg\}@scania.com}

\IEEEauthorblockA{\IEEEauthorrefmark{2}
{Department of Electrical And Information
Technology, LTH, Lund University, Sweden}\\
\{fr8354su-s, bi4215wi-s\}@student.lu.se}

}

\maketitle

\begingroup\renewcommand\thefootnote{\textsection}

\begin{abstract}
\textbf{[Context and Motivation]} The digital evolution of connected vehicles and the subsequent security risks emphasize the critical need for implementing in-vehicle cyber security measures such as intrusion detection and response systems to safeguard safety and functionality. 
The continuous advancement of attack scenarios further highlights the need for adaptive detection mechanisms that can detect evolving, unknown, and complex threats. The effective use of ML-driven techniques can help address this challenge. 
\textbf{[Problem]} However, constraints on implementing diverse attack scenarios on test vehicles due to safety, cost, and ethical considerations result in a scarcity of data representing attack scenarios. This limitation necessitates alternative efficient and effective methods for generating high-quality attack-representing data. \textbf{[Contribution]} This paper presents a context-aware attack data generator that generates attack inputs and corresponding in-vehicle network log, i.e., controller area network (CAN) log, representing various types of attack including denial of service (DoS), fuzzy, spoofing, suspension, and replay attacks. It utilizes parameterized attack models augmented with CAN message decoding and attack intensity adjustments to configure the attack scenarios with high similarity to real-world scenarios and promote variability. We evaluate the practicality of the generated attack-representing data within an intrusion detection system (IDS) case study, in which we develop and perform an empirical evaluation of two deep neural network IDS models using the generated data. In addition to the efficiency and scalability of the approach, the performance results of IDS models, high detection and classification capabilities, validate the consistency and effectiveness of the generated data as well. In this experience study, we also elaborate on the aspects influencing the fidelity of the data to real-world scenarios and provide insights into its application.



\end{abstract}



\section{Introduction}
The automotive industry is facing a significant transformation due to the digitalization of in-vehicle systems, advanced driver assistance systems (ADAS), and increasing vehicle connectivity as a backbone for advanced features. 
This digital evolution comes with inherent cyber security risks, as hackers may actively seek to manipulate vehicle software, access private data, and even use the vehicle for criminal actions, thereby posing threats to vehicle safety and consumer privacy. International regulations such as UN Regulation No 155 \cite{UN155}, mandate ensuring cyber security across the vehicle's lifecycle and require various measures to effectively detect and respond to cyber attacks on vehicles. 

The attacks may originate from multiple entry points including sensors, infotainment systems, telematics units, or direct physical interfaces. They can disrupt in-vehicle networks, interfere with data transmission, and cause electronic control units (ECUs) to malfunction. The controller area network (CAN) \cite{iso11898-1993} is a standard communication protocol widely used for data exchange between ECUs within a vehicle---which supports data rates of up to 1 Mbps with a payload size of 8 bytes. 
In 2012, CAN Flexible Data-rate (CAN-FD) was also introduced to enable higher data rates and larger payloads up to 64 bytes---standardized in ISO 11898-1:2015 \cite{iso11898-2015, iso11898-2024}. 
Despite several advantages of CAN, such as built-in error detection, cost-effectiveness due to reduced wiring, and a lightweight design, the CAN standard presents significant security vulnerabilities. The primary weaknesses include the absence of mechanisms for authentication, authorization, and encryption \cite{surveydeep}. These vulnerabilities result from the CAN standard's original design, which assumes all messages sent to an ECU are legitimate. Without authentication and authorization, malicious nodes can impersonate legitimate ECUs. Meanwhile, the lack of encryption means transparency of all CAN traffic, which can allow attackers to easily perceive and analyze data. Furthermore, as messages on the CAN bus are broadcast, a single compromised ECU could grant an attacker access to all transmitted information. In this context, the attackers might employ various disruptive mechanisms such as message fabrication (e.g., denial of service (DoS) and fuzzy attacks), suspension, masquerade (e.g, spoofing attack), and message replay \cite{UN155, comparative, roaddataset}.

One of the common security measures mandated by the regulation is the in-vehicle intrusion detection system (IDS), which runs on ECUs and provides a layer of security by analyzing the in-vehicle network traffic and identifying suspicious activities that could indicate an intrusion. The evolving threats and the continuous advancement of attack methods underscore the need for adaptive detection and response systems. In this context, data-driven solutions using ML techniques offer high potential to address this challenge by enabling the intrusion detection system to learn from the data representing the state of the vehicle, identify patterns, and adapt to new and emerging threats.

\textbf{Research problem.} The data representing the state of the vehicle under attack is \textit{scarce} as implementing all types of attack scenarios might not be feasible on the test vehicle, due to safety concerns, cost and resource limitations, and ethical considerations. Given these limitations, it is often required to consider effective alternatives such as synthetic generation of attack-representing data---to enrich the data required to develop ML-based security measures. Attack-representing data in this study refers to the in-vehicle network log, representing the state of the vehicle, under the occurrence of the attack.



It is worth noting that the need for (in-house) industry-driven research on enhancing the technical maturity of the in-vehicle security measures like in-vehicle (ECU-based) intrusion detection stems from the fact that relying on commercial products would imply (even partially) sharing sensitive information, e.g., network data and architecture-relevant information with the suppliers, which may be non-compliant with security requirements.
This industry experience paper intends to address the following research questions:\\
\textbf{RQ1}: How can we efficiently generate attack-representing CAN log data, covering various types of attack mechanisms such as fabrication, masquerade, suspension, and replay, based on normal CAN traffic?\\
\textbf{RQ2}: How practical is the generated attack-representing data, in terms of data quality characteristics such as consistency and effectiveness, for developing an intrusion detection security measure? \\
\textbf{RQ3}: What characteristics justify the fidelity of the generated data and what limitations remain?



This paper presents the following contributions:\\ 
\textbf{I.} We present a context-aware attack data generator that generates an attack-representing CAN log based on a given normal (attack-free) traffic dataset. It uses precise modeling of distinct attack classes and utilizes a set of adjustable parameters to configure the attack scenarios, which preserves the variability of attacks while retaining a realistic network flow. For frequency-disturbing attacks (i.e., DoS and fuzzy), in addition to adjustable parameters for timing of attack phases and generation of various fabricated messages, the attack modeling incorporates an attack intensity adjustment procedure as well. This involves adjusting the frequency of attack messages w.r.t the extracted patterns of an open-source real attack dataset, to mimic real attack intensity. For other classes of attacks based on impersonating, suspending, or replaying mechanisms, the attack modeling, besides adjusting the timing, also utilizes the CAN database (dbc) file to decode the raw messages, identify the communicating ECUs, and alter the data to accurately represent the intended attack scenarios.\\
\textbf{II.} To evaluate the practicality of the generated attack-representing data, we use it to develop two ML IDS models leveraging sequence-based deep neural networks (DNN), i.e., one- dimensional convolutional neural network (1D-CNN) and long short-term memory (LSTM) network, as a case study. We use a subset of the generated data for the model training and validation. We perform an empirical evaluation of the trained models  
on an unseen part of the data w.r.t various performance metrics---including the rate of false alarms and missed attack detection as two key performance indicators for IDS---and empirically show the consistency and effectiveness of the data. 

The high detection and classification capabilities of the IDS models developed using the generated data, i.e., false alarm and missed attack rates of $0.09 \%$ and $0.2 \%$ for the LSTM model; as well as $0.01 \%$ and $0.4 \%$ for the 1D-CNN, demonstrate high data consistency and effectiveness, as two quality characteristics of the data. 
Moreover, the efficiency (low cost) of the approach, the potential of scalability, understandability of the data due to the precise attack modeling all further highlight the practicality of the generated data. We also elaborate on the factors affecting the fidelity of the data to real-world scenarios and its potential application areas. We note that while parameterized attack models create scenarios closely resembling real-world attacks, the lack of observability of the actual effects of attacks on vehicle’s operation remains a limitation influencing the fidelity of the generated data.

The rest of the paper is organized as follows: Section \ref{Sec:Background} discusses background information on the structure of CAN messages and various types of attack models. Section \ref{Sec:Attack_Generator} presents the details of the attack data generator. Section \ref{Sec:IDS_Application} presents the IDS application developed based on the data. 
Section \ref{Sec:Results_Discussions} discusses and elaborates on the results answering the RQs 
and threats to validity of the results.  Section \ref{Sec:related_work} provides an overview of related work and insights into the applicability of the generated data and attack data generator, and Section~\ref{Sec:Conclusion} concludes the paper with a summary of the findings and research directions for future work.

\section{Background} \label{Sec:Background}

\subsection{CAN Frames}
The ISO 11898 series of standards presents the specification of the data-link and physical layers of the open systems interconnection (OSI) model for CAN communication \cite{iso11898-1993,iso11898-2024}. The data link layer sends messages using CAN frames, and it handles message framing, arbitration, error detection, and acknowledgment. 
The standard CAN frame uses an 11-bit identifier, while the extended format (CAN-FD) uses a 29-bit identifier, which is made up of the 11-bit base identifier plus an 18-bit extension. The CAN-FD frame supports both identifier formats and allows for a larger payload, up to 64 bytes \cite{iso11898-2024}. Vehicle manufacturers often keep the semantics of the CAN ID and data field proprietary and confidential.

The CAN protocol works based on a principle known as carrier-sense, multiple-access with collision resolution and arbitration on message priority (CSMA/CR+AMP) \cite{TexasIntro_CAN}. It means that each node on the network has to wait for the bus to be free before it can send a message, which is what CSMA means. 
The CR+AMP part means that if two messages are sent at the same time, there is a way to decide which one goes through. This happens through a process called bit-wise arbitration, where the message with higher priority, based on its identifier, will be allowed to continue. It prioritizes messages based on their identifiers in a way that a lower identifier has higher priority. The structure of a CAN frame, as shown in Fig. \ref{fig:CANframe}, includes the key fields, namely arbitration, control and data, cyclic redundancy check, and acknowledgment fields. The arbitration field contains the ID and the RTR/RRS bit (remote transmission request). Lower ID values mean higher priority and the RTR/RRS bit indicates whether the frame is a data frame carrying data or a remote frame, which is used to request data. The control field manages various data frame aspects such as data length code (DLC) indicating the size of the payload. The data field presents the message's payload. Cyclic redundancy check is used for transmission error detection and acknowledgment is used by the receivers to acknowledge that the data frame was received.  

\begin{figure}[h!]
    \centering
    \begin{subfigure}[b]{0.85\columnwidth}
        \centering
        \includegraphics[width=0.8\textwidth]{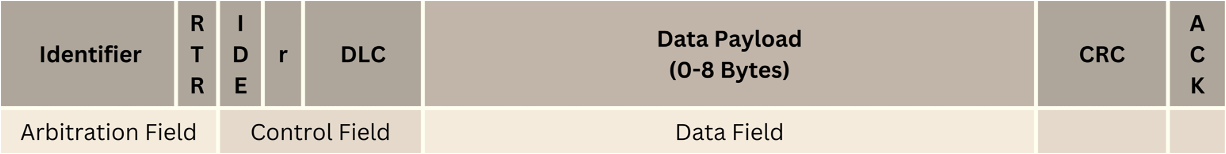}
        \caption{Standard CAN}
        \label{fig:frame}
    \end{subfigure}
    \begin{subfigure}[b]{\columnwidth}
        \centering
        \includegraphics[width=0.8\textwidth]{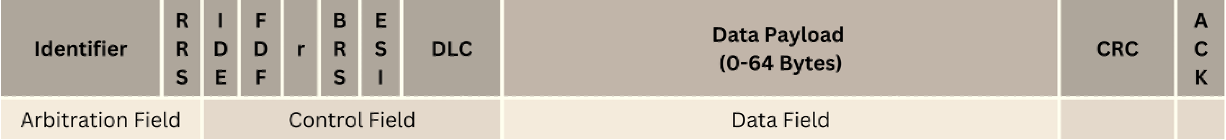}
        \caption{CAN-FD}
        \label{fig:fdframe}
    \end{subfigure}
    \caption{The CAN Frame}
    \label{fig:CANframe}
\end{figure}

\subsection{Attack Models}
\renewcommand{\thefootnote}{\roman{footnote}}
Attack surface analyses for connected vehicles, \cite{parkinson2017cyber, el2020cybersecurity, dadam2021onboard, pekaric2021taxonomy}, have investigated several potential attack vectors. An attack vector refers to a specific method or pathway that an attacker can use to gain unauthorized access to the system to inject harmful input or make malicious actions. It can be initiated through physical access such as via on-board diagnostic II (OBD-II port) interface or other surfaces, e.g., as reported in Kia boys challenge \cite{KiaBoys} and keyless RAV4 theft story\footnote{\url{https://kentindell.github.io/2023/04/03/can-injection/}}. Various sensors used in the vehicles 
could be exploited as attack entry points. Additionally, some of the vehicle systems with external communication interfaces such as the telematics module with remote commanding, infotainment, navigation, and remote key system, could be also among the direct target entry points \cite{parkinson2017cyber}.

Despite the range of attack vectors attackers may use, there are some core malicious mechanisms employed to disrupt normal flow on the CAN bus and have the connected systems malfunction. 
There are four representative categories of the attack mechanisms on the CAN bus \cite{fingerprintids, el2020cybersecurity}, i.e., fabrication, masquerade, suspension, and replay, that we consider and implement at least one type of attack for each mechanism.

\textbf{Fabrication.} An attacker, through gaining unauthorized access to the network and using an external device or through compromising an ECU, injects fabricated messages with falsified ID, DLC, and data---mainly aiming at the frequency disturbance. 
Flooding the bus or overriding messages from a legitimate ECU, are regarded as fabrication attacks, which can result in malfunctions or cause other ECUs to stop working. Fig. \ref{fig:Fabrication}, shows a simple example of a fabrication attack where a compromised ECU, C, sends multiple fabricated messages with falsified ID, which can block other messages with lower priority. 
DoS and fuzzy attacks are two types of attack based on the fabrication mechanism. 

\textit{Denial of Service (DoS):} The network is overwhelmed by a large number of messages, which can cause other ECUs to stop or delay their transmissions. The attacker often injects high-priority CAN messages, e.g., using ID 0x000, at a high frequency \cite{comparative, roaddataset}. 

\textit{Fuzzy attack:} It involves sending messages with random or semi-random IDs and data values. In this way, the attacker can inject malicious data into the network. 
By systematically testing the system with small fuzzy packets and observing how it reacts, the attacker can also gather information to develop further attacks \cite{securityanalysis}. 

\textbf{Masquerade.} 
The attacker sends forged malicious messages that seem to come from another ECU, while mimicking the normal message timing, to trick other ECUs into accepting them (e.g., keyless RAV4 theft case). It can be done involving two compromised ECUs: a weaker ECU that can be halted or suspended from sending messages, and a stronger ECU that the attacker controls to send malicious messages. 
The stealthy nature of this mechanism can cause considerable disruptions without changing the frequency of messages and pose a serious threat to the integrity of the network and vehicle safety. Fig. \ref{fig:Masquerade} shows a masquerade attack mechanism, illustrating that both ECU A (the weaker ECU) and ECU C (the stronger ECU) have been compromised. ECU A is prevented from transmitting any messages, while ECU C takes on the role of ECU A by sending malicious messages with ID A.

\textit{Spoofing attack:} It is a type of attack based on the masquerade mechanism. 
Here, the spoofing attack involves suspending a compromised ECU, creating and sending fake messages with a spoofed ID, so the attacker tricks other ECUs into accepting them as genuine. 

\textbf{Suspension.} The attacker compromises an ECU and prevents it from transmitting some or all of its messages. This disruption can cause malfunctions in the targeted ECU and in other ECUs that depend on its data. For instance, if the electric power steering ECU stops sending steering angle information, the stability control system, which relies on that data for traction control, will not function properly and can cause potentially dangerous situations \cite{fingerprintids}. Fig. \ref{fig:Suspension} demonstrates the suspension mechanism, where ECU C has been compromised and blocked from transmitting messages.

\textbf{Replay.} It involves capturing CAN messages for a certain period and retransmitting them at a later time. The attacker first monitors and records messages during the normal operation of the network. 
These recorded messages are then replayed to cause malfunctions. The replay attack uses unaltered messages, so it is often more difficult to detect. Fig. \ref{fig:Replay} shows the messages sent by ECU B are captured and replayed by compromised ECU C.

\begin{figure*}[h!]
    \centering
    \begin{subfigure}[b]{0.23\textwidth}
        \centering
        \includegraphics[width=\textwidth, height =3cm]{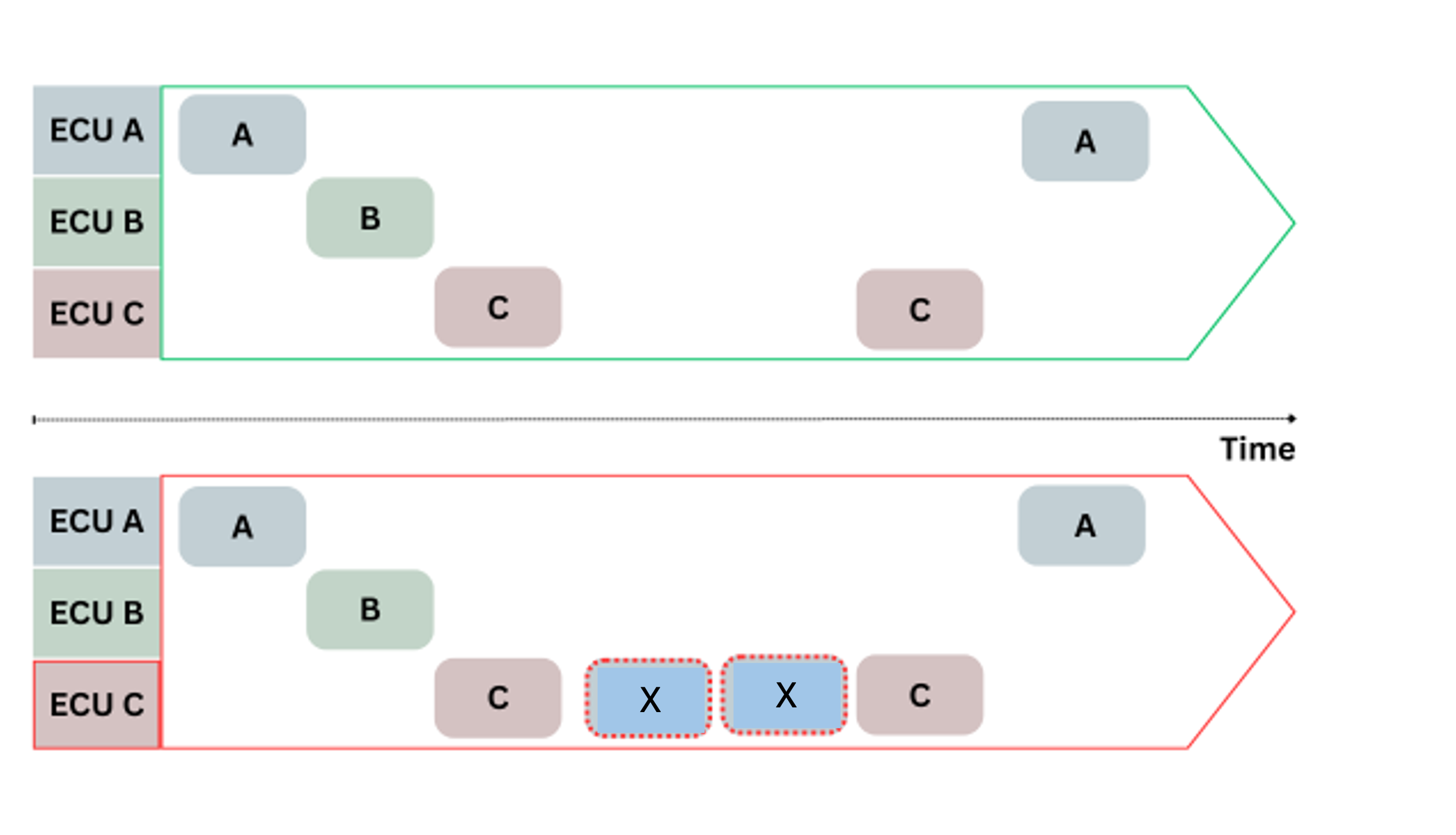}
        \caption{Fabrication}
        \label{fig:Fabrication}
    \end{subfigure}
    \begin{subfigure}[b]{0.23\textwidth}
        \centering
        \includegraphics[width=\textwidth, height =3cm]{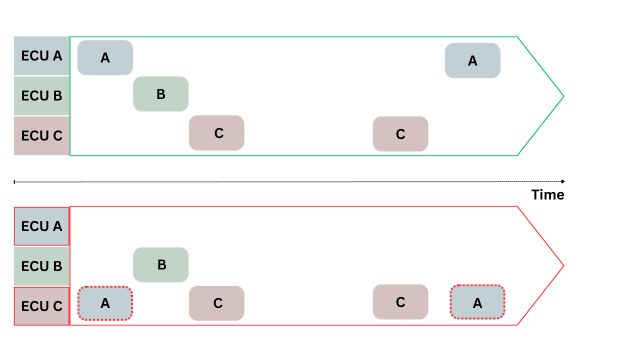}
        \caption{Masquerade}
        \label{fig:Masquerade}
    \end{subfigure}
    \begin{subfigure}[b]{0.23\textwidth}
        \centering
        \includegraphics[width=\textwidth, height =3cm]{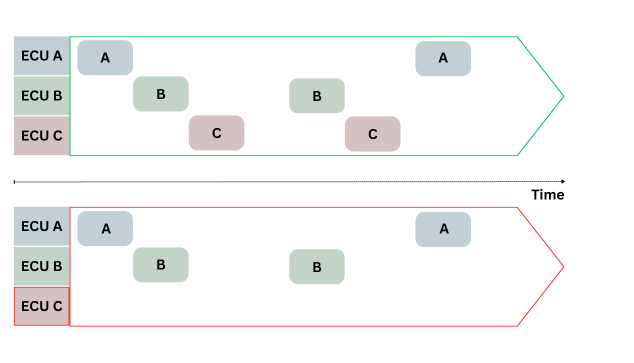}
        \caption{Suspension}
        \label{fig:Suspension}
    \end{subfigure}
    \begin{subfigure}[b]{0.23\textwidth}
        \centering
        \includegraphics[width=\textwidth, height =3cm]{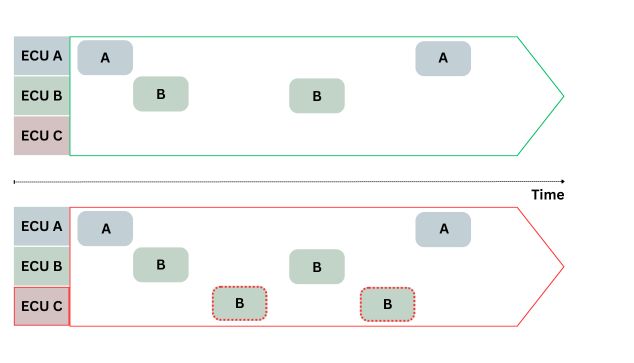}
        \caption{Replay}
        \label{fig:Replay}
    \end{subfigure}
    \caption{Attack mechanisms, the green arrow shows normal message flow and the red one shows the flow during attacks.}
    \label{fig:attack mechanism}
\end{figure*}

\section{Context-aware Attack Data Generation}\label{Sec:Attack_Generator}


\subsection{DoS and Fuzzy Attacks}

For generating the data representing frequency-disturbing attacks, such as DoS and fuzzy, the attack generator, first requires adjusting the frequency of attack messages to be inserted into the target normal log. To address this, it uses the extracted pattern from an open-source realistic CAN attack dataset, called Survival Analysis Dataset for automobile IDS \cite{survival_ids}. 

The open-source dataset contains CAN logs in both normal and under attack scenarios, collected via the OBD-II port from three different vehicle types---HYUNDAI YF Sonata, KIA Soul, and CHEVROLET Spark. The attack generator extracts the timing pattern from the open-source dataset, i.e., the time interval between the messages in the normal status, attack duration and frequency of attack messages in the attack mode. 
It extracts the timing pattern of the messages in the proprietary normal CAN log and using a scaling factor calculates the frequency of attack message injection so that an attack intensity same as the realistic open-source dataset will be mimicked and inserted into the CAN log.    

The proprietary and open-source datasets differ in vehicle types
as well as the open-source data consists of only standard CAN messages, whereas the proprietary dataset contains a mixed set of CAN and CAN-FD messages. Accordingly, a difference in message frequency is expected. The scale factor is utilized to consider the frequency difference and ensure that a similar attack intensity is created. Table \ref{tab:avg_intervals_normal} shows the average interval between all messages in the proprietary data and the open-source data. The scale factor is calculated from the average intervals, $\Bar{I}$, as shown in (\ref{scaling factor}). 

\begin{equation} \label{scaling factor}
\begin{split}
\text{Scale Factor} & = \frac{\Bar{I}_{\text{PROP.}}}{\frac{1}{3} ( \Bar{I}_{\text{KIA}} + \Bar{I}_{\text{SONATA}} + \Bar{I}_{\text{SPARK}} )} \\
&= \frac{0.0952}{\frac{1}{3}( 0.5121 + 0.4785 + 0.4351)}
\approx 0.2
\end{split}
\end{equation}

where $\Bar{I}_{\text{PROP.}}$, $\Bar{I}_{\text{KIA}}$, $\Bar{I}_{\text{SONATA}}$, and $\Bar{I}_{\text{SPARK}}$ indicate the average interval between messages in the normal (non-attack) proprietary, KIA, SONATA, and SPARK CAN logs, respectively.  

\begin{table}[b]
\centering
    \caption{Average interval between consecutive messages in the normal datasets}
    \label{tab:avg_intervals_normal}
\begin{tabular}{lllll}
\hline
Dataset       & Proprietary & SONATA & KIA & SPARK \\ \hline
Avg. interval (ms) &  0.0952 & 0.5121 & 0.4785 & 0.4351     \\ \hline
\end{tabular}
\end{table}

\textit{DoS Attack.} The Python script programmed to simulate a DoS attack generates a sequence of messages replicating a flood of CAN traffic and inserts it into the target log file. It creates bursts of high-priority messages (e.g., $ID = 0$) during specific attack phases, also known as
intrusion phases, with the aim of overwhelming the network.
The average interval between the attack messages, $D\Bar{I}_{\text{PROP.}}$, in the generated attack-representing log, is calculated as expressed in (\ref{attack frequency DoS}) (Table \ref{tab:avg_T_intervals}).

\begin{equation} \label{attack frequency DoS}
\begin{split}
D\Bar{I}_{\text{PROP.}} & = \frac{1}{3} ( D\Bar{I}_{\text{KIA}} + D\Bar{I}_{\text{SONATA}} + D\Bar{I}_{\text{SPARK}} ) \cdot 0.2 \\ & = \frac{1}{3} \cdot ( 0.6183 + 0.6059 + 0.6655  ) \cdot 0.2 \approx 0.13 \text { ms}
\end{split}
\end{equation}

where $D\Bar{I}_{\text{KIA}}$, $D\Bar{I}_{\text{SONATA}}$, $D\Bar{I}_{\text{SPARK}}$ indicate the average interval between DoS attack messages in the open-source dataset.  
The created attack phases are defined for random periods between 4 and 6 seconds, separated by pauses, to mimic the unpredictable nature of such attacks. The time interval between messages (with an average of approximately $0.13 ms$) and the pauses between bursts (with an average of $5 s$) are intentionally varied using a standard deviation to introduce variability into the timing. It means that the intervals and pauses fluctuate around these averages and it makes the timing more realistic or less predictable.


\textit{Fuzzy Attack.} The generator selects a set of the most frequent message IDs from  the proprietary normal data and uses it to create pseudo-random IDs through an XOR operation along with a random number---to enhance the diversity of IDs while preserving the general format of valid messages. Each generated ID has a random payload to simulate message content.
These fabricated messages are inserted into proprietary normal log file during the attack phases---with are defined in periods of $4s$ to $6s$ alongside the pause periods with an average of $5s$. The average interval between the fuzzy attack messages, $F\Bar{I}_{\text{PROP.}}$, is approximately $0.2 ms$ (\ref{attack frequency fuzzy}). Similar to DoS scenario, the interval between the attack messages and pauses between the attack phases both undergo a standard deviation to increase the variability and reduce predictability.  

\begin{equation} \label{attack frequency fuzzy}
\begin{split}
F\Bar{I}_{\text{PROP.}} & = \frac{1}{3}( F\Bar{I}_{\text{KIA}} + F\Bar{I}_{\text{SONATA}} + F\Bar{I}_{\text{SPARK}}) \cdot 0.2  \\ & = \frac{1}{3} \cdot ( 0.9746 + 1.0142 + 0.9334  ) \cdot 0.2 \approx 0.2 \text { ms}
\end{split}
\end{equation}

where $F\Bar{I}_{\text{KIA}}$, $F\Bar{I}_{\text{SONATA}}$, and $F\Bar{I}_{\text{SPARK}}$ indicate the interval between the fuzzy attack messages in the open-source dataset.
Table \ref{table:DoS_fuzzy_attack_parameters} summarizes the adjustable parameters to configure DoS and fuzzy attack scenarios. 
Regarding the integration of fabricated messages into the target log data, timestamps of the messages are examined and if needed, are adjusted to prevent collisions and preserve the arbitration ID-based prioritization---to represent a realistic network flow.

\begin{table}[b]
\centering
    \caption{Average interval between attack messages in fuzzy and DoS}
    \label{tab:avg_T_intervals}
\begin{tabular}{lllll}
\hline
      & Proprietary & SONATA & KIA    & SPARK  \\ \hline
DoS   & 0.1260          & 0.6183 & 0.6059 & 0.6655 \\
Fuzzy & 0.1948          & 1.0142 & 0.9746 & 0.9334 \\ \hline
\end{tabular}
\end{table}

\begin{table}[b]
\centering
\caption{DoS and fuzzy attack configuration parameters}
\label{table:DoS_fuzzy_attack_parameters}
\begin{tabular}{|p{0.3\columnwidth}|p{0.3\columnwidth}|}
\hline
\textbf{Parameter}          & \textbf{Value}                                  \\ \hline
Attack phases     & 4 to 6 s                               \\ \hline
Average of pause period          & 5 s                                        \\ \hline
Average of attack message interval        & $D\Bar{I}_{\text{PROP.}} = 0.13ms$,\newline $F\Bar{I}_{\text{PROP.}} = 0.2ms$                               \\ \hline
Standard deviation for pauses & ${\text{Average of Pause}}/{10}$   \\ \hline
Standard deviation for message intervals &  ${D\Bar{I}_{\text{PROP.}}}/{3}$, \newline  ${F\Bar{I}_{\text{PROP.}}}/{3}$ \\ \hline
Arbitration ID selection    & \textit{DoS}: 0, \newline  \textit{Fuzzy}: Pseudo-random IDs based on top 33\% frequent IDs                                                \\ \hline
Payload                     & Randomly generated data                         \\ \hline
\end{tabular}
\end{table}

\subsection{Spoofing Attack}
The spoofing attack generator script works by replacing legitimate messages from a specific ECU with new ones containing pseudo-random payloads---while keeping the same message IDs, which preserves the normal transmission cycle of the ECU. It creates pseudo-random payloads based on the actual data from the target ECU's messages by applying XOR operation with a random number on the original payload data---similarly to the pseudo-random IDs in the fuzzy attack. A standard deviation is also applied to the message timestamps to simulate timing variations. The generator script extracts the messages originating from the target ECU by encoding the messages with a CAN database (dbc file) and mapping the ECU to the message IDs, so that it ensures that only messages of the target ECU are considered for spoofing. 
So, unlike the fuzzy and DoS attacks, which operate in separate attack phases, the spoofing attack occurs continuously over a period.

\begin{table}[h]
\centering
\caption{Spoofing attack configuration parameters}
\label{table:Spoofing_attack_parameters}
\begin{tabular}{|p{0.3\columnwidth}|p{0.3\columnwidth}|}
\hline
\textbf{Parameter}                       & \textbf{Value}                                                                 \\ \hline                                            
Attack messages' behavior               & Same as the cyclic behavior of normal messages                                  \\ \hline
Standard deviation for timestamps of attack messages  & 1\% of the base (original) cyclic time difference                                              \\ \hline
Arbitration ID selection                      & Uses the same IDs as the targeted ECU's messages                               \\ \hline
Payload                    & Pseudo-random based on original payload                \\ \hline
\end{tabular}
\end{table}

\subsection{Suspension Attack}
The attack generator script replicates the effect of temporarily suspending a target ECU from the network by selectively removing messages sent by the target ECU over a set period. It uses the dbc files to decode the messages and by using the ECU-to-message ID mapping it identifies the message IDs sent by the target ECU. Once identified, these messages are filtered out from the CAN log during the specified time window. It creates a gap in the target CAN log that mimics the ECU being suspended from the network.

\subsection{Replay Attack}
To generate data representing replay attack, it targets messages with frequently occurring IDs in normal CAN log---the top 5 most frequent IDs---and replays them according to their respective cyclic timing. To preserve the consistency and integrity of the network flow, the average interval between the target messages is calculated. Then, the target messages are duplicated, and their timestamps are adjusted so that their cyclic timing is preserved. These messages are integrated into the CAN log over a set period. Fig. \ref{fig:overview_generator} and Algorithm \ref{alg:generate_attack_data} show an overview of the attack data generation process.

\begin{figure}[h]
    \centering
\includegraphics[width=0.75\columnwidth, height=5.5cm]{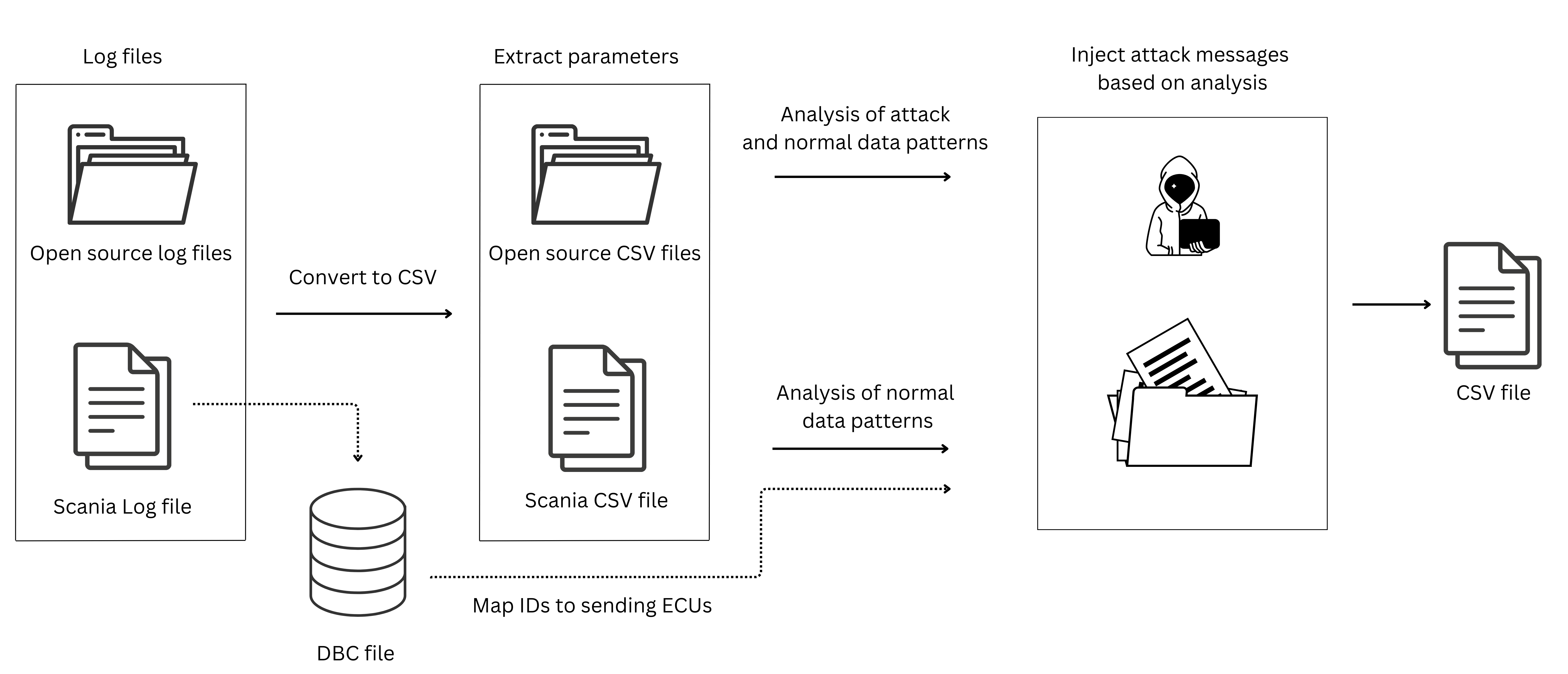}
    \caption{Attack generation process}
    \label{fig:overview_generator}
\end{figure}

\begin{algorithm}[h]
\caption{GenerateAttackData}
\label{alg:generate_attack_data}
\KwIn{proprietaryLog, attackType, configParams}
\KwOut{attack-representingLog}
\If{attackType == "DoS" OR attackType == "Fuzzy"}{
    pattern $\gets$ ExtractMsgInterval(srcData, proprietaryLog)\;
    scale $\gets$ CalcScaleFactor(proprietaryLog, srcData)\;
    adjFreq $\gets$ AdjustAttackMsgFreq(pattern, scale)\;
    attackMsgs $\gets$ CreateFabricatedMsgs(adjFreq)\;
    IntegrateMsgs(proprietaryLog, attackMsgs, configParams.DoSFuzzyAtkPhase)\;
}
\If{attackType == "Spoofing"}{
    targetECU $\gets$ IdentifyECU(proprietaryLog, dbcFile)\;
    spoofMsgs $\gets$ ReplacePayload(tgtECU, proprietaryLog)\;
    IntegrateMsgs(proprietaryLog, spoofMsgs)\;
}
\If{attackType == "Suspension"}{
    tgtECU $\gets$ IdentifyECU(proprietaryLog, dbcFile)\;
    suspendedLog $\gets$ RemoveMsgs(proprietaryLog, tgtECU, configparams.suspensionWindow)\;
}
\If{attackType == "Replay"}{
    freqIDs $\gets$ ExtractIDs(proprietaryLog, configparams.topN)\;
    replayMsgs $\gets$ DuplicateMsgs(proprietaryLog, freqIDs)\;
    IntegrateMsgs(proprietaryLog, replayMsgs)\;
}

\Return{attack-representingLog}\;
\end{algorithm}



\section{A Data-driven Security Measure: IDS Application} \label{Sec:IDS_Application}
An IDS application, to detect and identify different classes of attacks, is developed utilizing two widely used DNN models, LSTM and 1D-CNN, and empirically evaluated based on an unseen part of the generated data. 
LSTM is effective for handling time-series data by capturing sequential patterns and CNN is noted for its feature extraction capabilities to identify distinct patterns. 

\subsection{Data Management} \label{Sec:Data_Management}
To prepare training and testing data for the models, 3 million messages over the same time span (period) were selected from each attack-contained data file, i.e., DoS-, fuzzy-, spoofing-, and replay-representing attack CAN log, and a dataset containing 12 million messages is considered in this case study. In the CAN logs, each entry consists of attributes of timestamp, CAN ID, DLC, payload, and type. The 'type' is labeled 'T' for the attack messages and 'R' for normal ones. 


To utilize the sequence processing of the DNN models, the data 
is organized into message sequences using a sliding window and stride value set to $1$. Two sequence lengths of $25$ and $50$ are selected empirically. The sequences are labeled as 0 (Normal) if there are no attack messages within that sequence and 1 (DoS), 2 (Fuzzy), 3 (Spoofing), or 4 (Replay) if there is at least one attack message within the sequence. The percentage of the normal and attack sequences in the selected dataset is shown in Table \ref{table:benign_attack_Sequences}. 
The data is split into training, validation, and testing data, as 70\% is used for training, 15\% for validation, and the remaining 15\% for testing. 

\begin{table}[b]
\centering
\caption{Normal and attack instances}
\label{table:benign_attack_Sequences}
\begin{tabular} {@{}p{1.5cm}|p{2.5cm}|p{2.5cm}@{}}
\toprule
\textbf{} & \multicolumn{2}{c}{Number of Seq. (\% of total) } \\
\hline
Class   &  $length = 25$  & $length = 50$ \\ \midrule
Normal   &  5,586,608 (46.6\%) & 4,244,333 (36.4\%) \\
DoS      &  1,895,101   (15.8\%) & 1,895,601 (15.8\%)   \\
Fuzzy    &  1,810,856 (15.1\%) &  1,811,406 (15.1\%)\\
Spoofing &  847,344 (7.1\%) & 1,510,206 (12.6\%)\\
Replay   &  1,859,995 (15.5\%) & 2,538,258 (21.2\%)\\ \bottomrule
\end{tabular}
\end{table}

\subsection{Feature Selection and Model Architecture}
The selected features are derived from the CAN ID and the first eight bytes of payload data. The CAN ID is converted from hexadecimal to a 29-bit binary format, with each bit used as a distinct feature. Each payload byte (initially in hexadecimal) is transformed to decimal, then normalized, and regarded as one feature. The feature selection resulting in 37 features aims to balance computational efficiency for training while preserving sufficient information. It is done empirically w.r.t the domain knowledge and considering factors like the prevalence of standard CAN messages over CAN-FD in the data log and corresponding target segments of the in-vehicle network, and findings from related studies.

The LSTM model comprises two LSTM layers---each layer having 50 units---followed by a Dense layer with 50 units using sigmoid activation, and an output Dense layer with 5 units using softmax activation. The 1D-CNN model features an input layer accepting the input sequences, followed by a 1D convolutional layer with 256 filters, a kernel size of 3, and sigmoid activation. The convolutional output is then downsampled with a MaxPooling1D layer before being flattened. Then, it includes a Dense layer with 64 units and sigmoid activation, followed by an output layer with 5 units and softmax activation for the multi-class classification. The models are compiled with the Nadam optimizer (learning rate of 0.0001) and use categorical cross-entropy as the loss function. A batch size of 1024 and 256 is chosen for training the LSTM and CNN models respectively, and both models are trained over 100 epochs.

\section{Results and Discussion} \label{Sec:Results_Discussions}

\subsection{RQ1: Attack-Representing Data Generation}
\textit{DoS and fuzzy.} Figures \ref{fig:DoS_plots} and \ref{fig:Fuzzy_plots} show the scatter plot of messages---both normal and attack messages---in generated propriety attack-representing and open-source datasets, for DoS and fuzzy respectively. Attack messages, labeled T, are shown along with normal messages, labeled R, and each red section highlights an attack interval. For proprietary data, only a 100-second time span is plotted to provide a comparative view with the total duration of data in the open-source dataset, which is quite smaller. The x-axis denotes each message's timestamp, and the y-axis shows the interval between consecutive messages. The dotted line represents the average message interval under attack. 


\textit{Spoofing.} Fig. \ref{Spoofing_under_attack} presents the plot of the messages over a time span in which a targeted ECU is under spoofing attack and  
for comparison, Fig. \ref{normal_targeted_ECU} shows the messages over the same time span in a normal state. 

\textit{Suspension.} Fig. \ref{fig:Suspension_plot} shows the messages transmitted by a targeted ECU during a suspension attack episode. The gap shown in the plot represents a 200-second suspension period, in which no messages were sent by the targeted ECU. 

\begin{figure*}[h]
    \centering
    \begin{subfigure}[b]{0.49\textwidth}
        \centering
        \includegraphics[width=0.95\textwidth, height=4.5cm]{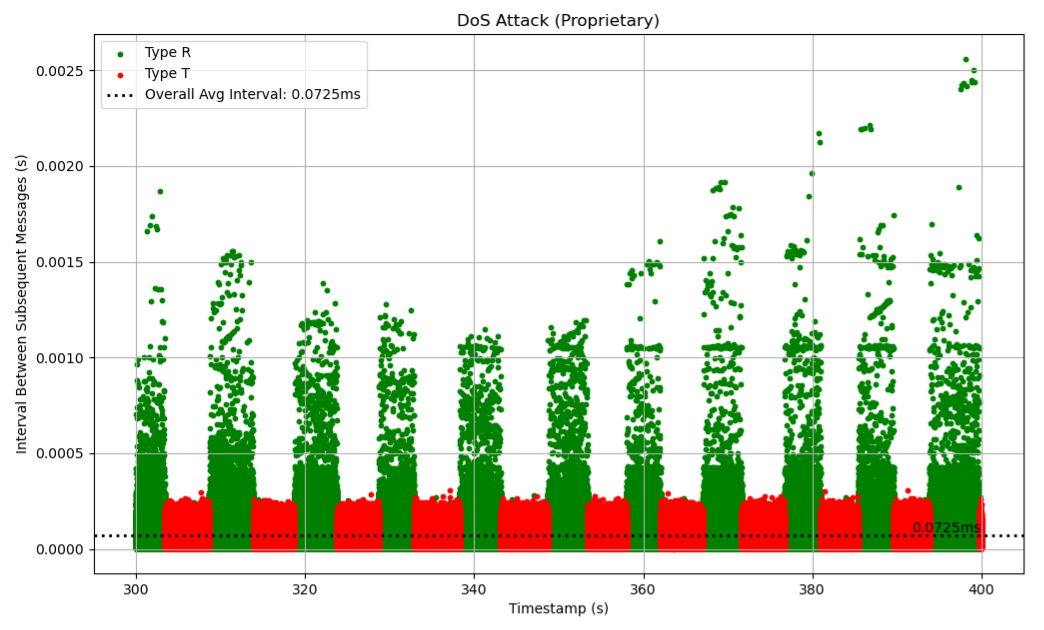}
        \caption{Proprietary data}
    \end{subfigure}
    \begin{subfigure}[b]{0.49\textwidth}
        \centering
        \includegraphics[width=0.95\textwidth, height=4.5cm]{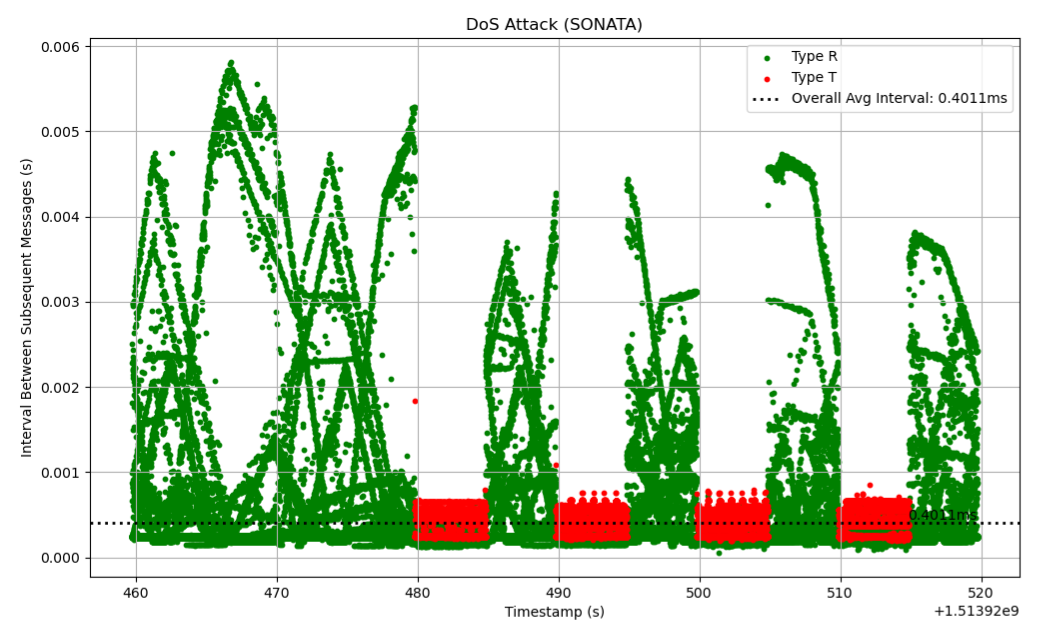}
        \caption{SONATA data}
    \end{subfigure}
    \newline 
    \begin{subfigure}[b]{0.49\textwidth}
        \centering
        \includegraphics[width=0.95\textwidth, height=4.5cm]{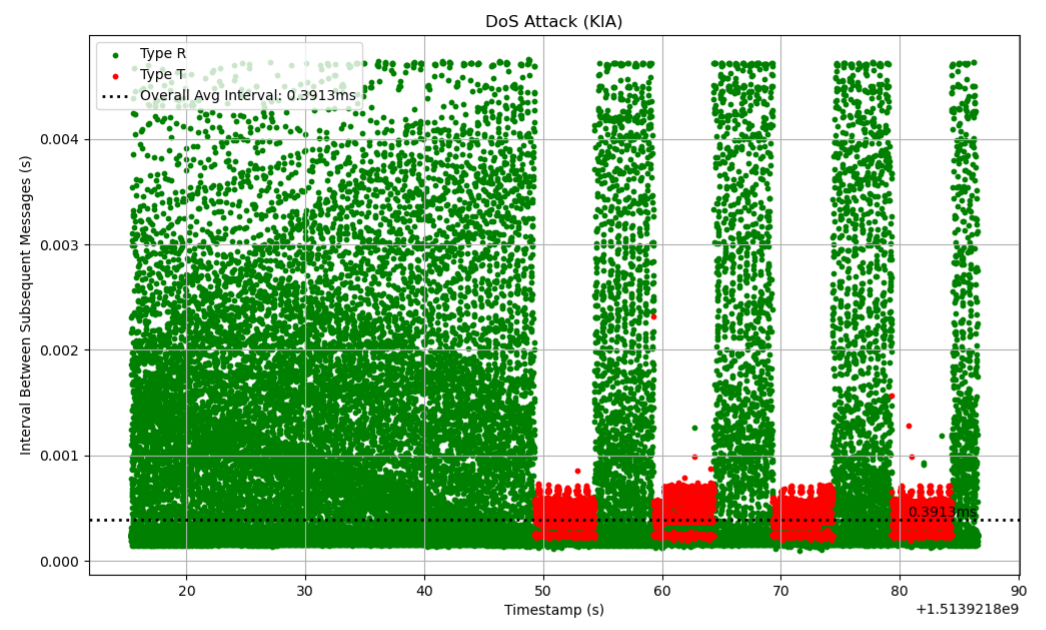}
        \caption{KIA data}
    \end{subfigure}
    \begin{subfigure}[b]{0.49\textwidth}
        \centering
        \includegraphics[width=0.95\textwidth, height=4.5cm]{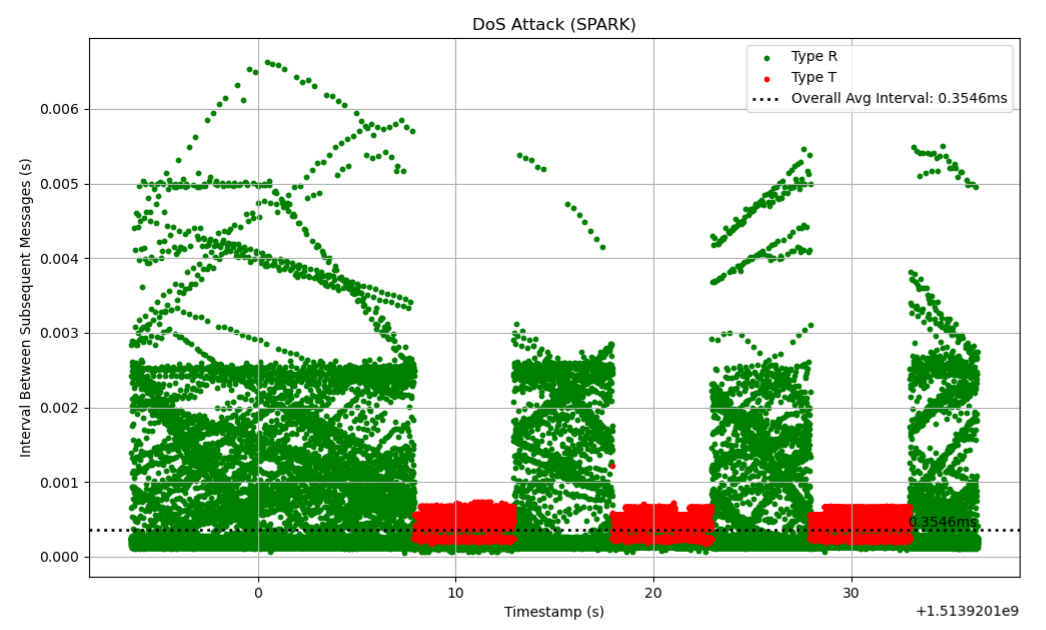}
        \caption{SPARK data}
    \end{subfigure}
    \caption{DoS: The scatter plot of the messages under DoS attack. Attack and normal messages are marked in red and green respectively. The horizontal axis shows the timestamps, and the vertical axis indicates the interval between consecutive messages.}
    \label{fig:DoS_plots}
\end{figure*}

\textit{Replay.} Fig. \ref{fig:Replay_zoomed_plot} illustrates a snapshot of the network during a replay attack, which indicates the replay of messages. The top five most frequent messages were targeted for replay. In the plot, green crosses show normal messages, while red crosses mark those same messages that are replayed.\\ 

\tikzstyle{mybox} = [draw=black, very thick,
    rectangle, rounded corners, inner sep=5pt, inner ysep=5pt]

\begin{tikzpicture}
\node [mybox] (box){%
    \begin{minipage}{0.95\columnwidth}
        RQ1: By utilizing parameterized attack models augmented with attack intensity adjustment and message decoding, we propose a cost-efficient approach to generate attack-representing CAN log based on a given normal CAN traffic, without the need for test vehicles.
    \end{minipage}
};
\end{tikzpicture}

\subsection{RQ2: Data Practicality in the IDS Case Study }
To show the practicality of the generated data, we perform an empirical evaluation of the IDS models using an unseen part of the generated data (test data) and show how \textit{effective} and \textit{consistent} the data is. 
Effectiveness of the data refers to its quality to meet the requirements for the intended ML task, which is the attack detection and classification in our case study. This characteristic ensures that the dataset is adequate quantitatively and qualitatively to support the training of a model delivering acceptable performance. Consistency mainly refers to data being without conflicting records and errors. Consistency is also essential for ML models to ensure robust and well-performing learning \cite{ISO/IEC5259-2, mowla2024guide}.

\begin{figure*}[t]
    \centering
    \begin{subfigure}[b]{0.49\textwidth}
        \centering
        \includegraphics[width=0.95\textwidth, height=4.5cm]{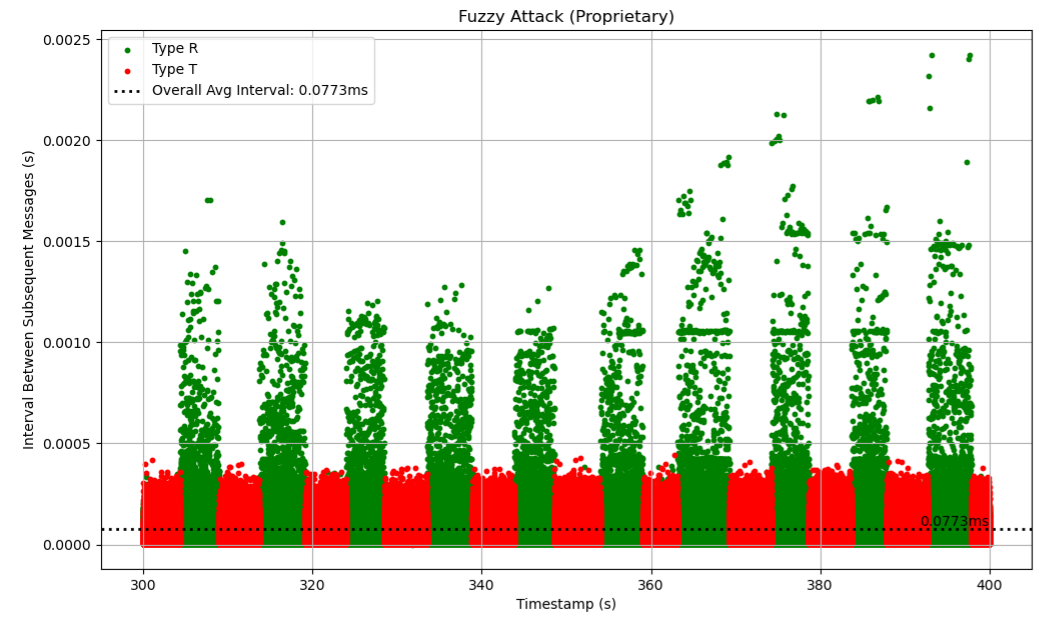}
        \caption{Proprietary data}
    \end{subfigure}
    \begin{subfigure}[b]{0.49\textwidth}
        \centering
            \includegraphics[width=0.95\textwidth, height=4.5cm]{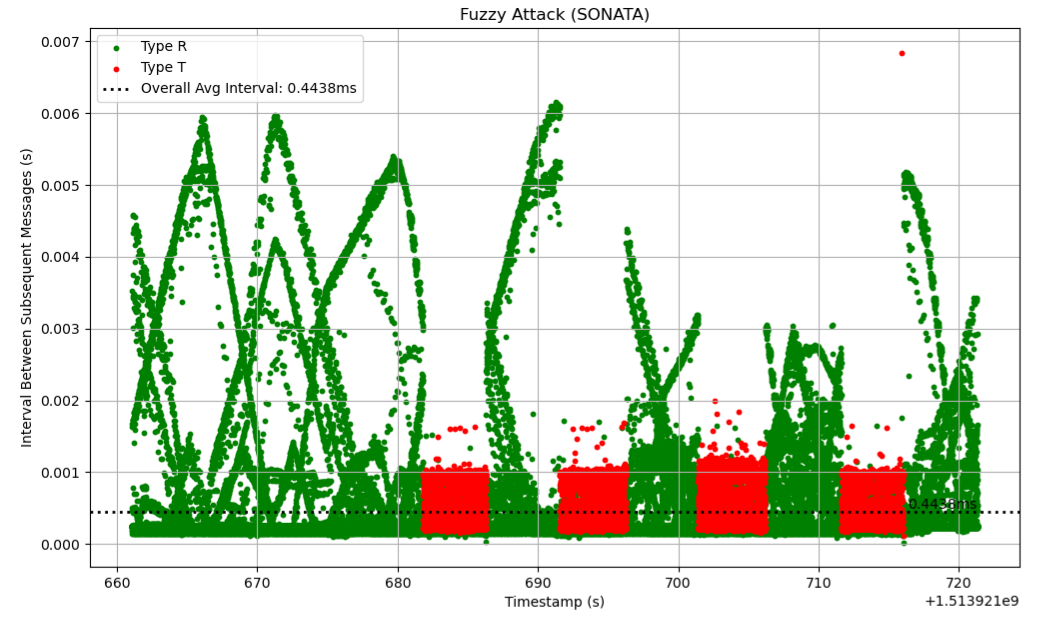}
        \caption{SONATA data}
    \end{subfigure}
    \newline 
    \begin{subfigure}[b]{0.49\textwidth}
        \centering
        \includegraphics[width=0.95\textwidth, height=4.5cm]{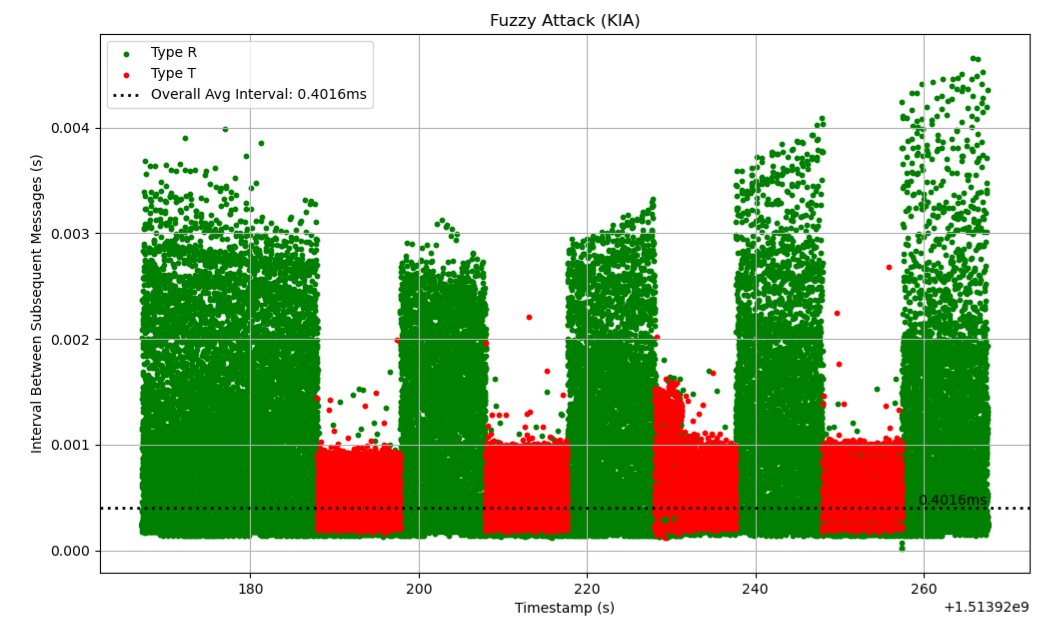}
        \caption{KIA data}
    \end{subfigure}
    \begin{subfigure}[b]{0.49\textwidth}
        \centering
        \includegraphics[width=0.95\textwidth, height=4.5cm]{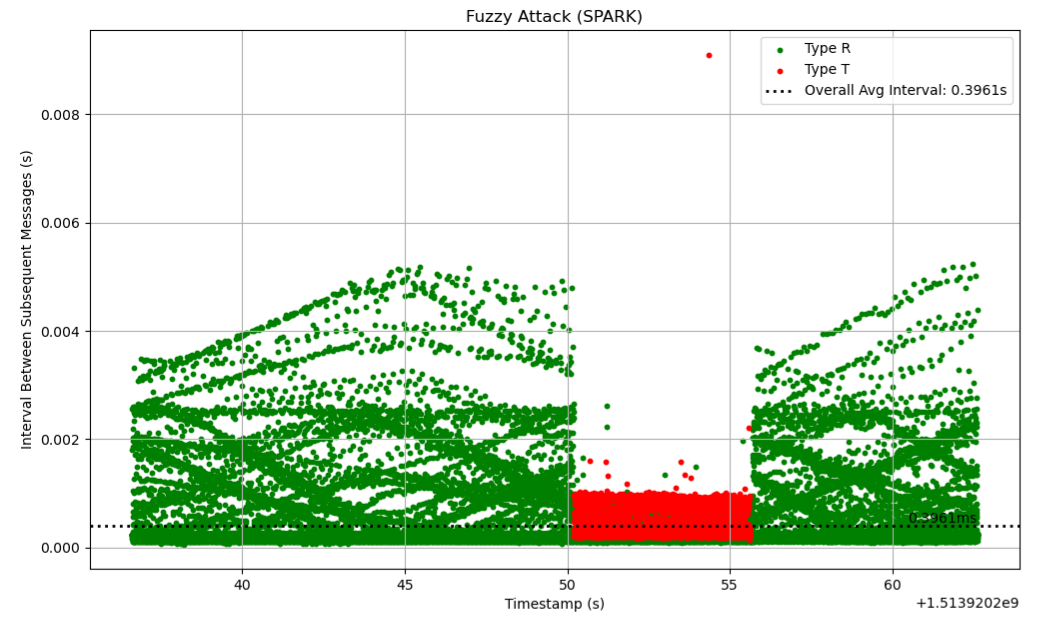}
        \caption{SPARK data}
    \end{subfigure}
    \caption{Fuzzy: The scatter plot of the messages under fuzzy attack.}
    \label{fig:Fuzzy_plots}
\end{figure*}

\begin{figure*}[h]
    \centering\begin{subfigure}[b]{0.49\textwidth}
        \centering
        \includegraphics[width=0.95\textwidth, height=4.5cm]{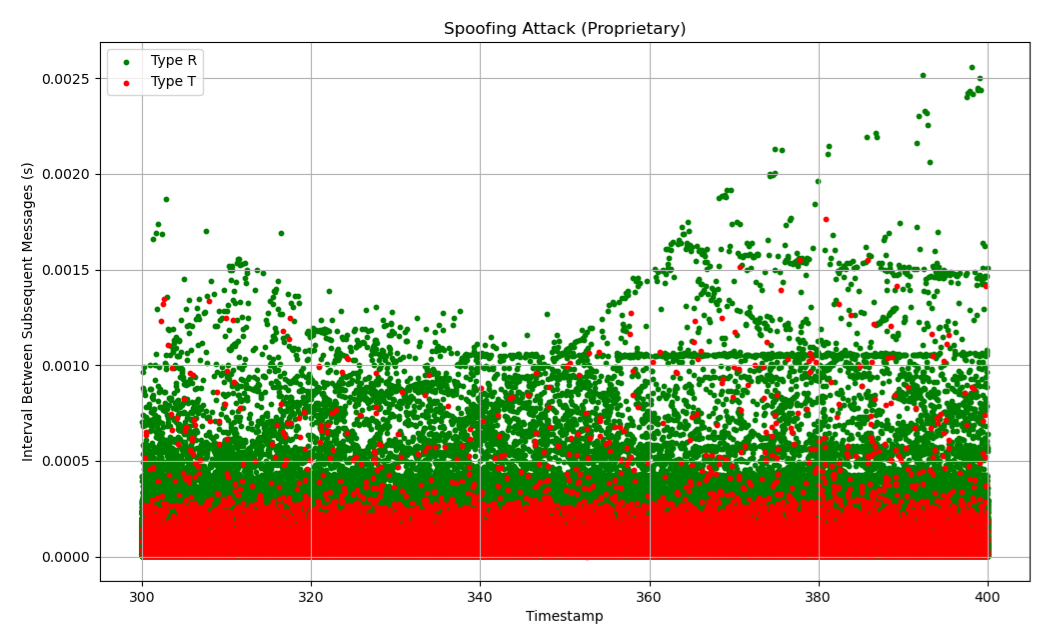}
        \caption{Under spoofing attack}
        \label{Spoofing_under_attack}
    \end{subfigure}
    \begin{subfigure}[b]{0.49\textwidth}
        \centering
        \includegraphics[width=0.95\textwidth, height=4.5cm]{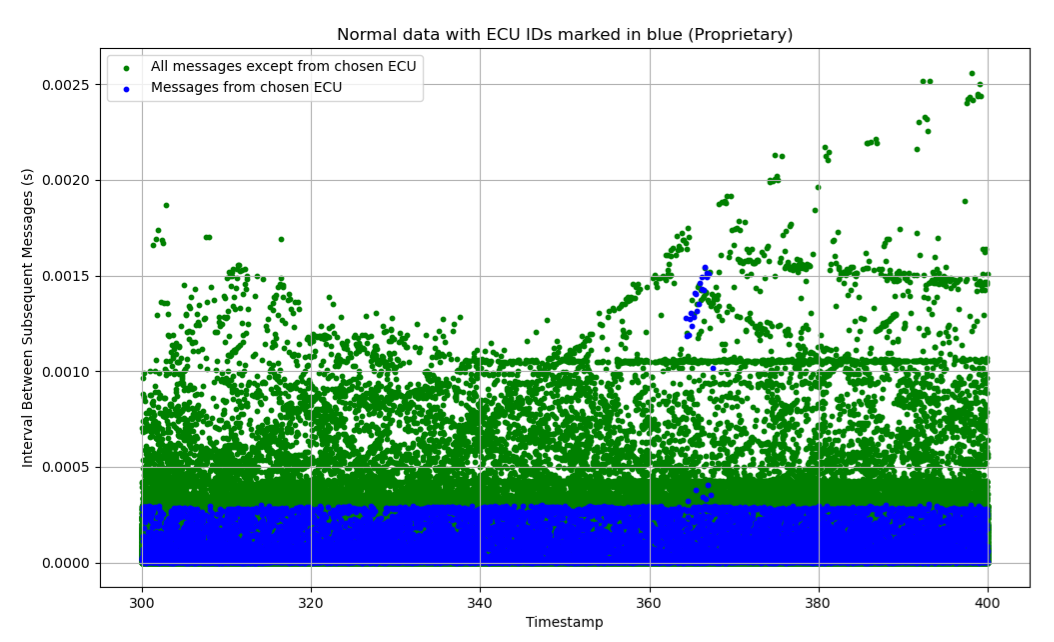}
        \caption{Normal state}
        \label{normal_targeted_ECU}
    \end{subfigure}
    \caption{Spoofing: The scatter plot of the messages over a time span, w.r.t a specific target ECU. Attack messages (T) are marked in red, original messages from the targeted ECU in blue, and other normal messages (R) are marked in green.}
    \label{fig:Spoofing_plot}
\end{figure*}

\begin{figure}[h]
    \centering
\includegraphics[width=0.55\columnwidth, height=4.7cm]{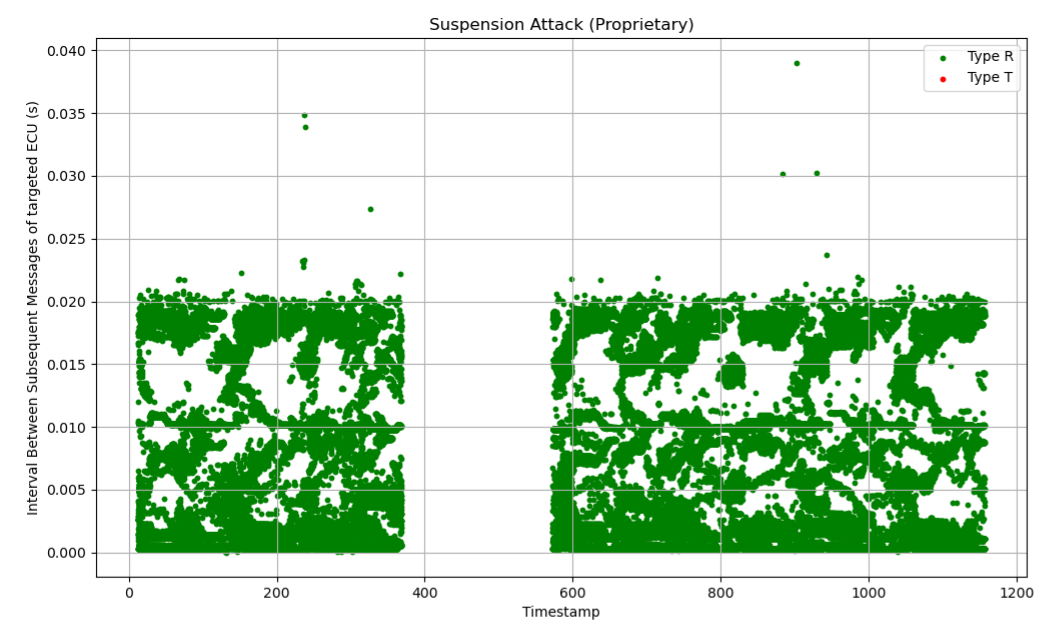}
    \caption{Suspension: The scatter plot of the messages over a time span in which a targeted ECU is under suspension attack.}
    \label{fig:Suspension_plot}
\end{figure}

\begin{figure}[h]
    \centering
\includegraphics[width=0.55\columnwidth, height= 4.5cm]{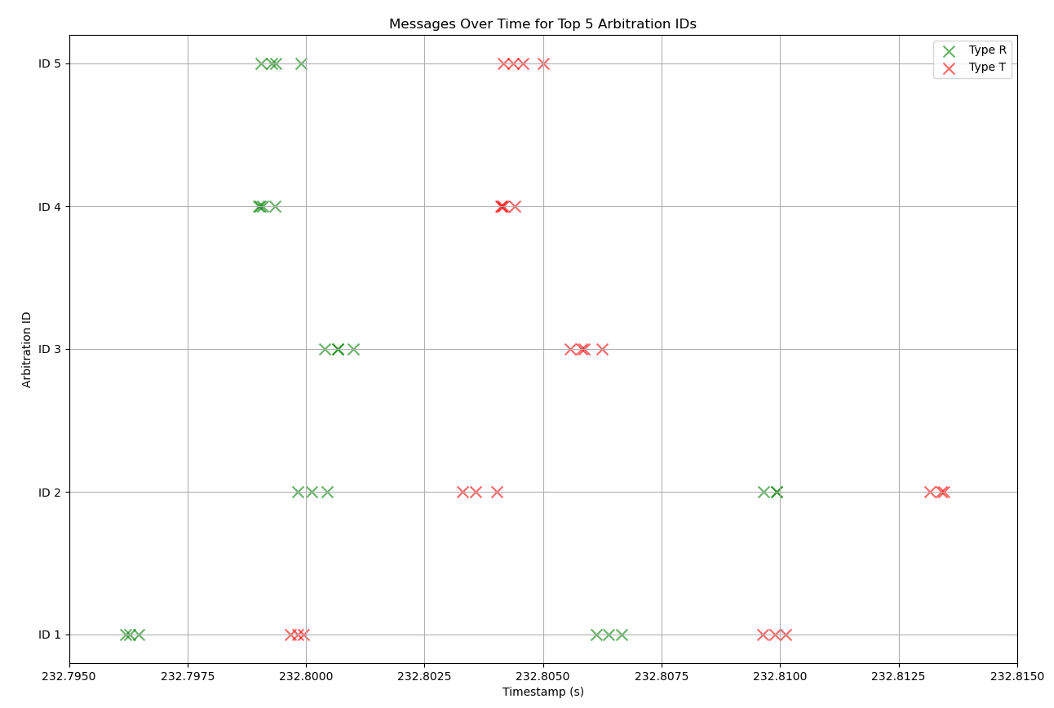}
    \caption{Normal and replayed messages during a time span under replay attack.}
    \label{fig:Replay_zoomed_plot}
\end{figure}


The following performance metrics are considered in the evaluation of the IDS models. Detection accuracy (\ref{eq:detection_accuracy}), indicates the overall correctness of the model's classifications across all classes. Precision (\ref{eq:precision}), 
indicates the proportion of true positive predictions for a target class, $c$ . 
Recall (\ref{eq:recall}), measures the model's sensitivity or its ability to correctly identify all actual instances of the target class. The F1 score (\ref{eq:f1_score}), provides a balanced metric, which harmonizes precision and recall. 
The False Positive Rate (FPR) (\ref{eq:false_positive_rate}), reflects the fraction of negative instances of the target class that were misclassified as positive. 
False Negative Rate (FNR) (\ref{eq:false_negative_rate}), measures the proportion of positive instances incorrectly classified as negative.

\begin{equation}
Detection\ Accuracy = \frac{TP + TN}{TP + TN + FP + FN}
\label{eq:detection_accuracy}
\end{equation}

\begin{equation} \label{eq:precision}
Precision_c = \frac{TP_c}{TP_c + FP_c}
\end{equation}

\begin{equation} \label{eq:recall}
Recall_c = \frac{TP_c}{TP_c + FN_c}
\end{equation}

\begin{equation} \label{eq:f1_score}
F1\ Score_c = 2 \times \frac{Precision_c \times Recall_c}{Precision_c + Recall_c}
\end{equation}

\begin{equation}
FPR_c = \frac{FP_c}{TN_c + FP_c}
\label{eq:false_positive_rate}
\end{equation}

\begin{equation}
FNR_c = \frac{FN_c}{TP_c + FN_c}
\label{eq:false_negative_rate}
\end{equation}


Table \ref{table:lstm_results_25} and Table \ref{table:lstm_50_results} summarize the performance evaluation of the LSTM models trained on the input sequences of length 25 and 50. Then, Tables \ref{table:cnn_results_1D_25} and \ref{table:cnn_results_1D_50} present the performance of the 1D-CNN models trained on input sequences of lengths 25 and 50, respectively. 
In addition to the overall accuracy, two key performance indicators highly considered are the FNR and the FPR for the Normal class, which indicate the normal messages that are misclassified as attacks (false alarms) and the actual attack messages that are incorrectly classified as normal, respectively. Besides the comparable high accuracy of the LSTM and 1D-CNN models, both classes of models show high performance regarding the rate of false alarms and missed attacks, in particular with longer input sequences---learning based on a longer episode of the messages. \\


\begin{tikzpicture}
\node [mybox] (box){%
    \begin{minipage}{0.95\columnwidth}
        RQ2: The performance results of the IDS models developed using the generated data, demonstrate  the effectiveness and consistency of the data in supporting training, ensuring reliable learning, and avoiding overfitting. This proves the practicality of the generated data for the development of in-vehicle IDS measures.
    \end{minipage}
};
\end{tikzpicture}

\begin{table}[t]
\centering
\caption{Performance of IDS LSTM model, input sequences of length 25 }
\label{table:lstm_results_25}
\begin{tabular}{@{}p{0.85cm}|p{1cm}|p{1cm}|p{0.8cm}|p{0.8cm}|p{0.8cm}|p{0.8cm}@{}}
\toprule
\textbf{Class} & \textbf{Accuracy} & \textbf{Precision} & \textbf{Recall} & \textbf{F1 Score} & \textbf{FPR} & \textbf{FNR} \\ \midrule
Normal  & \textbf{} & 0.9928 & 0.9963 & 0.9945 & \textbf{\underline{0.0063}} & \textbf{\underline{0.0037}} \\
DoS     & \textbf{} & 1.0000 & 0.9999 & 1.0000 & 0.0000 & 0.0001 \\
Fuzzy   & {99.49 \%} & 0.9999 & 0.9994 & 0.9997 & 0.0000 & 0.0006 \\
Spoofing & \textbf{} & 0.9968 & 0.9839 & 0.9903 & 0.0002 & 0.0161 \\
Replay  & \textbf{} & 0.9899 & 0.9861 & 0.9880 & 0.0018 & 0.0139 \\ \bottomrule
\end{tabular}
\end{table}

\begin{table}[t]
\centering
\caption{Performance of IDS LSTM model, input sequences of length 50)}
\label{table:lstm_50_results}
\begin{tabular}{@{}p{0.85cm}|p{1cm}|p{1cm}|p{0.8cm}|p{0.8cm}|p{0.8cm}|p{0.8cm}@{}}
\toprule
\textbf{Class} & \textbf{Accuracy} & \textbf{Precision} & \textbf{Recall} & \textbf{F1 Score} & \textbf{FPR} & \textbf{FNR} \\ \midrule
Normal  & \textbf{} & 0.9965 & 0.9991 & 0.9978 & \textbf{\underline{0.0019}} & \textbf{\underline{0.0009}} \\
DoS     & \textbf{} & 1.0000 & 0.9999 & 1.0000 & 0.0000 & 0.0001 \\
Fuzzy   & {99.84 \%} & 1.0000 & 0.9999 & 0.9999 & 0.0000 & 0.0001 \\
Spoofing & \textbf{} & 0.9998 & 0.9981 & 0.9990 & 0.0000 & 0.0019 \\
Replay  & \textbf{} & 0.9986 & 0.9953 & 0.9970 & 0.0004 & 0.0047 \\ \bottomrule
\end{tabular}
\end{table}

\begin{table}[t]
\centering
\caption{Performance of IDS 1D-CNN model, input sequences of length 25}
\label{table:cnn_results_1D_25}
\begin{tabular}{@{}p{0.85cm}|p{1cm}|p{1cm}|p{0.8cm}|p{0.8cm}|p{0.8cm}|p{0.8cm}@{}}
\toprule
\textbf{Attack} & \textbf{Accuracy} & \textbf{Precision} & \textbf{Recall} & \textbf{F1 Score} & \textbf{FPR} & \textbf{FNR} \\ \midrule
Normal  & \textbf{} & 0.9916 & 0.9964 & 0.9940 & \textbf{\underline{0.0074}} & \textbf{\underline{0.0036}}\\
DoS     & \textbf{} & 1.0000 & 1.0000 & 1.0000 & 0.0000 & 0.0000 \\
Fuzzy   & {99.44 \%} & 0.9998 & 0.9998 & 0.9998 & 0.0000 & 0.0002 \\
Spoofing & \textbf{} & 0.9997 & 0.9761 & 0.9877 & 0.0000 & 0.024 \\
Replay  & \textbf{} & 0.9895 & 0.9856 & 0.9876 & 0.0019 & 0.0144 \\ \bottomrule
\end{tabular}
\end{table}

\begin{table}[t]
\centering
\caption{Performance of IDS 1D-CNN model, input sequences of length 50}
\label{table:cnn_results_1D_50}
\begin{tabular}{@{}p{0.85cm}|p{1cm}|p{1cm}|p{0.8cm}|p{0.8cm}|p{0.8cm}|p{0.8cm}@{}}
\toprule
\textbf{Attack} & \textbf{Accuracy} & \textbf{Precision} & \textbf{Recall} & \textbf{F1 Score} & \textbf{FPR} & \textbf{FNR} \\ \midrule
Normal  & \textbf{} & 0.9924 & 0.9999 & 0.9961 & \textbf{\underline{0.0042}} & \textbf{\underline{0.0001}} \\
DoS     & \textbf{} & 1.0000 & 1.0000 & 1.0000 & 0.0000 & 0.0000 \\
Fuzzy   & {99.72 \%} & 1.0000 & 0.9999 & 0.9999 & 0.0000 & 0.0001 \\
Spoofing & \textbf{} & 0.9999 & 0.9991 & 0.9995 & 0.0000 & 0.0009 \\
Replay  & \textbf{} & 0.9998 & 0.9878 & 0.9938 & 0.0000 & 0.0122 \\ \bottomrule
\end{tabular}
\end{table}

\subsection{RQ3: Fidelity Analysis of the Generated Attack-Representing Data}
One of the main hindrances in the fidelity assessment of the generated attack-contained data is the lack of observability of the effects of attacks on the vehicle’s operation. Unlike implementing attacks in real mode, where the impacts on the systems are observable, synthetic attacks lack the realism of presenting the impacts. 


\textit{Similarity to real scenarios.} The fabrication attack data (DoS and fuzzy) is benchmarked compared to an open-source real dataset, which demonstrates a similar average ratio of attack messages in attack phases, as shown in Table \ref{tab:avg_ratio}. 
The average percentage of attack messages within an attack phase, i.e., the number of attack messages over the total number of messages, in the generated attack-representing data, is aligned with the open-source real dataset. Specifically, for the DoS attack, the proprietary data shows the ratio of attack messages of 38.6\%, and the ratio of attack messages in the open-source dataset also ranges from 35.6\% to 39.3\%. For the fuzzy attack, the proportion is approximately 28\% across all.

For a suspension attack, in which the main feature is the removal of messages, synthetic attack data can closely resemble the real-world scenario by reflecting the absence of messages from the target ECUs. Similarly, for replay attacks, 
with a proper mechanism for preserving the integrity of the timestamps, the  
synthetic data can keep high fidelity to real-world scenarios. Regarding the spoofing attack, achieving full realism can be more challenging due to the nuances of timing and content, which are difficult to precisely replicate. \\


\begin{tikzpicture}
\node [mybox] (box){%
    \begin{minipage}{0.95\columnwidth}
        RQ3: The parameterized attack modeling and attack intensity adjustments utilized in the proposed approach enhance the similarity of the generated attacks to the real-world scenarios, however the lack of observability of the actual effects of the attacks remains a limitation.
    \end{minipage}
};
\end{tikzpicture}

\section{Related Work} \label{Sec:related_work}
Various approaches can be utilized to generate attack data for in-vehicle CAN traffic. One approach involves using HiL and SiL environments along with simulation tools that can simulate the in-vehicle networks such as CANoe \cite{CANoe}. 
ATG (Attack Traffic Generation) \cite{huang2018atg} is an open-source vehicle CAN bus analyzer working with USB2CAN bus interface. It supports message sniffing, message replay, and fabricated message generation and it can be used as a cheaper option than commercial tools like CANoe in some cases. Neelap and Bhandari \cite{trafficgenerator} explore a synthetic attack data generation approach utilizing analysis of CAN logs with a focus on the replay and fuzzy attacks---altering the messages without context interpretation. Toyama et al. \cite{toyama2018pasta} introduce a portable automotive security testbed (called PASTA), which supports monitoring in-vehicle CAN buses and provides the possibility to sniff CAN buses and inject random fabricated messages.  

The development of CAN IDS has been the focus of considerable research work by both researchers and practitioners \cite{rathore2022vehicle, lokman2019intrusion}. Designing an IDS solution involves balancing multiple trade-offs between accuracy, latency, efficiency, and scalability; and each class of IDS solutions exhibits unique strengths and limitations in these aspects. 
The ML-based solutions are intended to distinguish normal communication patterns from abnormal ones, using datasets that either strictly contain normal data or contain a combination of normal and abnormal patterns. The effective use of ML-driven techniques can help address the challenge of evolving, unknown, and complex threats stemming from the changing nature of cyber attacks. They can provide adaptability to evolving threats, detection of unknown patterns, and behavioral analysis of the networks.
The ML-based models employed for CAN IDS include various forms of deep learning models such as LSTM-based classification \cite{hossain2020lstm}, language models \cite{alkhatib2022can}, and autoencoders \cite{hoang2022detecting}.  
Regarding the resource constraints on the classical vehicle ECUs, an interest in the development of lightweight models \cite{basavaraj2022towards} as well as hybrid models \cite{zhang2022hybrid}, which combine models to balance accuracy and computation cost w.r.t the available resources is also seen in the recent work.

\textbf{Insights on the application.} The attack data generator offers high efficiency (cost), and usability \cite{strandberg2023westermo}. It can also be used with or integrated into the simulation environments (e.g., CANoe) with the capability of log replaying. As shown in the case study, the generated attack-representing data can be efficiently used to build ML-driven in-vehicle IDS. When a limited set of real attack-contained data is available, the generated attack data can be used to build pre-trained IDS models, which can be further fine-tuned with the real attack-contained data. The generated attack-representing data is also highly useful for adjusting the anomaly detection threshold in semi-supervised IDS models---trained only on normal data.

\section{Conclusion} \label{Sec:Conclusion}
This study presents a context-aware attack data generator enabling the automated generation of attack-contained CAN log representing diverse attack scenarios, i.e., DoS, fuzzy, spoofing, suspension, and replay attacks, without the need for physical test vehicles. It provides an efficient and effective alternative for implementing real attack scenarios on the test vehicles.
It can facilitate the development of ML-driven security measures such as IDS applications. We demonstrate its practicality and show the effectiveness and consistency of the generated attack-representing data within an IDS case study---in which we develop and perform an empirical evaluation of two DNN-based IDS models using the generated data.  
Concerning other data quality characteristics, besides the efficiency (the low cost) of the proposed approach and its support for scalable attack data generation, utilizing the dbc files and message decoding ensures strong alignment with the vehicle system and enhances the data understandability. Lastly, the attack models considered in the approach align with the requirements of the regulations, e.g., Reg. No 155 \cite{UN155}. Meanwhile, the generated attack-representing data can be used in various ways for developing ML-driven in-vehicle intrusion detection systems.
Future works will include using LLMs i.e., GPT and Llama, based on few-shot prompting to further enrich the attack-contained data and further facilitate the development of intelligent in-vehicle intrusion detection security measures.

\begin{table}[H]
\centering
    \caption{Ave. ratio of attack messages in DoS and fuzzy}
    \label{tab:avg_ratio}
\begin{tabular}{lllll}
\hline
      & Proprietary & SONATA & KIA    & SPARK  \\ \hline
DoS (\%)  & 38.610      & 39.269 & 35.645 & 35.645 \\
Fuzzy (\%) & 28.293      & 28.565 & 28.170 & 28.170 \\ \hline
\end{tabular}
\end{table}

\section*{Acknowledgment}
This work has been supported by the SCANIA internal research project PreSuccess and by Vinnova through the research project INTERSTICE (INTelligent sEcuRity SoluTIons for Connected vEhicles), under grant number 2024-00661.

\bibliographystyle{IEEEtran}
\bibliography{reference}

\begin{thebibliography}{10}
\providecommand{\url}[1]{#1}
\csname url@samestyle\endcsname
\providecommand{\newblock}{\relax}
\providecommand{\bibinfo}[2]{#2}
\providecommand{\BIBentrySTDinterwordspacing}{\spaceskip=0pt\relax}
\providecommand{\BIBentryALTinterwordstretchfactor}{4}
\providecommand{\BIBentryALTinterwordspacing}{\spaceskip=\fontdimen2\font plus
\BIBentryALTinterwordstretchfactor\fontdimen3\font minus \fontdimen4\font\relax}
\providecommand{\BIBforeignlanguage}[2]{{%
\expandafter\ifx\csname l@#1\endcsname\relax
\typeout{** WARNING: IEEEtran.bst: No hyphenation pattern has been}%
\typeout{** loaded for the language `#1'. Using the pattern for}%
\typeout{** the default language instead.}%
\else
\language=\csname l@#1\endcsname
\fi
#2}}
\providecommand{\BIBdecl}{\relax}
\BIBdecl

\bibitem{UN155}
\BIBentryALTinterwordspacing
{The United Nations Economic Commission for Europe (UNECE)}, ``{UN Regulation No. 155},'' 2023. [Online]. Available: \url{{https://unece.org/transport/documents/2021/03/standards/un-regulation-no-155-cyber-security-and-cyber-security}}
\BIBentrySTDinterwordspacing

\bibitem{iso11898-1993}
\BIBentryALTinterwordspacing
{International Organization for Standardization}, ``{ISO 11898: Road vehicles -- Interchange of digital information -- Controller area network (CAN) for high-speed communication},'' Geneva, Switzerland, 1993. [Online]. Available: \url{https://www.iso.org/standard/20380.html}
\BIBentrySTDinterwordspacing

\bibitem{iso11898-2015}
\BIBentryALTinterwordspacing
------, ``{ISO 11898-1:2015 Road vehicles -- Controller area network (CAN) -- Part 1: Data link layer and physical signalling},'' Geneva, Switzerland, 2015. [Online]. Available: \url{https://www.iso.org/standard/63648.html}
\BIBentrySTDinterwordspacing

\bibitem{iso11898-2024}
\BIBentryALTinterwordspacing
------, ``{ISO 11898: Road vehicles -- Road vehicles — Controller area network (CAN) -- Part 1: Data link layer and physical coding sublayer},'' Geneva, Switzerland, 2024. [Online]. Available: \url{https://www.iso.org/standard/86384.html}
\BIBentrySTDinterwordspacing

\bibitem{surveydeep}
B.~Lampe and W.~Meng, ``A survey of deep learning-based intrusion detection in automotive applications,'' \emph{Expert Systems with Applications}, vol. 221, 2023, doi: \href{https://doi.org/10.1016/j.eswa.2023.119771}{10.1016/j.eswa.2023.119771}.

\bibitem{comparative}
B.~S. Bari, K.~Yelamarthi, and S.~Ghafoor, ``{Intrusion Detection in Vehicle Controller Area Network (CAN) Bus Using Machine Learning: A Comparative Performance Study},'' \emph{Sensors}, vol.~23, no.~7, p. 3610, 2023, doi: \href{https://doi.org/10.3390/s23073610}{10.3390/s23073610}.

\bibitem{roaddataset}
M.~E. Verma, R.~A. Bridges, M.~D. Iannacone, S.~C. Hollifield, P.~Moriano, S.~C. Hespeler, B.~Kay, and F.~L. Combs, ``{A comprehensive guide to CAN IDS data and introduction of the ROAD dataset},'' \emph{PLoS ONE}, vol.~19, no.~1, pp. 1 -- 32, 2024, doi: \href{https://doi.org/10.1371/journal.pone.0296879}{10.1371/journal.pone.0296879}.

\bibitem{TexasIntro_CAN}
\BIBentryALTinterwordspacing
{TEXAS INSTRUMENTS}, ``{Introduction to the Controller Area Network (CAN)},'' Texas, USA, 2016. [Online]. Available: \url{{https://www.ti.com/lit/an/sloa101b/sloa101b.pdf}}
\BIBentrySTDinterwordspacing

\bibitem{parkinson2017cyber}
S.~Parkinson, P.~Ward, K.~Wilson, and J.~Miller, ``Cyber threats facing autonomous and connected vehicles: Future challenges,'' \emph{IEEE transactions on intelligent transportation systems}, vol.~18, no.~11, pp. 2898--2915, 2017.

\bibitem{el2020cybersecurity}
Z.~El-Rewini, K.~Sadatsharan, D.~F. Selvaraj, S.~J. Plathottam, and P.~Ranganathan, ``Cybersecurity challenges in vehicular communications,'' \emph{Vehicular Communications}, vol.~23, p. 100214, 2020.

\bibitem{dadam2021onboard}
S.~R. Dadam, D.~Zhu, V.~Kumar, V.~Ravi, and V.~S.~S. Palukuru, ``Onboard cybersecurity diagnostic system for connected vehicles,'' SAE Technical Paper, Tech. Rep., 2021.

\bibitem{pekaric2021taxonomy}
I.~Pekaric, C.~Sauerwein, S.~Haselwanter, and M.~Felderer, ``A taxonomy of attack mechanisms in the automotive domain,'' \emph{Computer Standards \& Interfaces}, vol.~78, p. 103539, 2021.

\bibitem{KiaBoys}
\BIBentryALTinterwordspacing
{The Verge}, ``{The Kia Boys will steal your car for clout},'' 2023. [Online]. Available: \url{{https://www.theverge.com/23742425/kia-boys-car-theft-steal-tiktok-hyundai-usb}}
\BIBentrySTDinterwordspacing

\bibitem{fingerprintids}
K.-T. Cho and K.~G. Shin, ``{Fingerprinting electronic control units for vehicle intrusion detection},'' in \emph{Proceedings of the 25th USENIX Conference on Security Symposium}.\hskip 1em plus 0.5em minus 0.4em\relax USA: USENIX Association, 2016, p. 911–927, doi: \href{https://dl.acm.org/doi/10.5555/3241094.3241165}{10.5555/3241094.3241165}.

\bibitem{securityanalysis}
K.~Koscher, A.~Czeskis, F.~Roesner, S.~Patel, T.~Kohno, S.~Checkoway, D.~McCoy, B.~Kantor, D.~Anderson, H.~Shacham, and S.~Savage, ``{Experimental Security Analysis of a Modern Automobile},'' Oakland, CA, USA, 2010, pp. 447 -- 462, doi: \href{https://ieeexplore.ieee.org/document/5504804}{10.1109/SP.2010.34}.

\bibitem{survival_ids}
\BIBentryALTinterwordspacing
{Hacking and Countermeasure Research Lab (HCRL)}, ``{Survival Analysis Dataset for automobile IDS},'' 2019. [Online]. Available: \url{{https://ocslab.hksecurity.net/Datasets/survival-ids}}
\BIBentrySTDinterwordspacing

\bibitem{ISO/IEC5259-2}
\BIBentryALTinterwordspacing
{ISO/IEC}, ``{ISO/IEC5259-2 Artificial intelligence — Data quality for analytics and machine learning (ML)},'' 2024. [Online]. Available: \url{{https://www.iso.org/standard/81860.html}}
\BIBentrySTDinterwordspacing

\bibitem{mowla2024guide}
N.~I. Mowla, ``{A Guide to Data Quality Testing for AI Applications based on Standards, RISE Research Institutes of Sweden},'' 2024.

\bibitem{CANoe}
\BIBentryALTinterwordspacing
{Vector}, ``{Development and Test Tools for Automotive HIL and SIL Projects},'' 2023. [Online]. Available: \url{{https://www.vector.com/int/en/products/products-a-z/software/canoe }}
\BIBentrySTDinterwordspacing

\bibitem{huang2018atg}
T.~Huang, J.~Zhou, and A.~Bytes, ``{ATG: An Attack Traffic Generation Tool for Security Testing of In-vehicle CAN Bus},'' in \emph{Proceedings of the 13th International Conference on Availability, Reliability and Security}, no.~32.\hskip 1em plus 0.5em minus 0.4em\relax Association for Computing Machinery, 2018, p.~6, doi: \href{https://doi.org/10.1145/3230833.3230843}{10.1145/3230833.3230843}.

\bibitem{trafficgenerator}
C.~Neelap and H.~V. Bhandari, ``{Attack Traffic Generation for Network-based Intrusion Detection System},'' M.S. Thesis, Chalmers University of Technology and University of Gothenburg, Gothenburg, Sweden, 2023.

\bibitem{toyama2018pasta}
T.~Toyama, T.~Yoshida, H.~Oguma, and T.~Matsumoto, ``Pasta: Portable automotive security testbed with adaptability,'' \emph{London, blackhat Europe}, 2018.

\bibitem{rathore2022vehicle}
R.~S. Rathore, C.~Hewage, O.~Kaiwartya, and J.~Lloret, ``In-vehicle communication cyber security: challenges and solutions,'' \emph{Sensors}, vol.~22, no.~17, p. 6679, 2022.

\bibitem{lokman2019intrusion}
S.-F. Lokman, A.~T. Othman, and M.-H. Abu-Bakar, ``{Intrusion detection system for automotive Controller Area Network (CAN) bus system: a review},'' \emph{EURASIP Journal on Wireless Communications and Networking}, no. 184, pp. 1--17, 2019, doi: \href{https://jwcn-eurasipjournals.springeropen.com/articles/10.1186/s13638-019-1484-3}{10.1186/s13638-019-1484-3}.

\bibitem{hossain2020lstm}
M.~D. Hossain, H.~Inoue, H.~Ochiai, D.~Fall, and Y.~Kadobayashi, ``Lstm-based intrusion detection system for in-vehicle can bus communications,'' \emph{Ieee Access}, vol.~8, pp. 185\,489--185\,502, 2020.

\bibitem{alkhatib2022can}
N.~Alkhatib, M.~Mushtaq, H.~Ghauch, and J.-L. Danger, ``Can-bert do it? controller area network intrusion detection system based on bert language model,'' in \emph{2022 IEEE/ACS 19th International Conference on Computer Systems and Applications (AICCSA)}.\hskip 1em plus 0.5em minus 0.4em\relax IEEE, 2022, pp. 1--8.

\bibitem{hoang2022detecting}
T.-N. Hoang and D.~Kim, ``Detecting in-vehicle intrusion via semi-supervised learning-based convolutional adversarial autoencoders,'' \emph{Vehicular Communications}, vol.~38, p. 100520, 2022.

\bibitem{basavaraj2022towards}
D.~Basavaraj and S.~Tayeb, ``Towards a lightweight intrusion detection framework for in-vehicle networks,'' \emph{Journal of Sensor and Actuator Networks}, vol.~11, no.~1, p.~6, 2022.

\bibitem{zhang2022hybrid}
L.~Zhang and D.~Ma, ``A hybrid approach toward efficient and accurate intrusion detection for in-vehicle networks,'' \emph{IEEE Access}, vol.~10, pp. 10\,852--10\,866, 2022.

\bibitem{strandberg2023westermo}
P.~E. Strandberg, D.~S{\"o}derman, A.~Dehlaghi-Ghadim, M.~Leon, T.~Markovic, S.~Punnekkat, M.~H. Moghadam, and D.~Buffoni, ``The westermo network traffic data set,'' \emph{Data in Brief}, vol.~50, p. 109512, 2023.

\end{thebibliography}

\end{document}